\documentstyle[preprint,aps]{revtex}

\begin{document}
\draft

\preprint{US-98-06/AMU-98-6-1}

\title{
Universal Seesaw Mass Matrix Model\\
with Three Light Pseudo-Dirac Neutrinos
}

\author{\bf Yoshio Koide\thanks{E-mail: koide@u-shizuoka-ken.ac.jp}}
\address{Department of Physics, University of Shizuoka \\ 
52-1 Yada, Shizuoka 422-8526, Japan}

\author{\bf Hideo Fusaoka\thanks{E-mail: fusaoka@amugw.aichi-med-u.ac.jp}}
\address{Department of Physics, Aichi Medical University \\ 
Nagakute, Aichi 480-1195, Japan}


\maketitle

\begin{abstract}
A universal seesaw mass matrix model, which
gives successful description of quark mass matrix 
in terms of lepton masses, yields three ``sterile" neutrinos 
$\nu^s_{i}$, which compose pseudo-Dirac neutrinos 
$\nu_{i\pm}^{ps}\simeq (\nu_{i} \pm \nu^s_{i})/\sqrt{2}$ 
together with the active neutrinos $\nu_{i}$
($i=e,\mu,\tau$). 
The solar and atmospheric neutrino data are
explained by the mixings 
$\nu_{e}\leftrightarrow \nu^s_{e}$ and 
$\nu_{\mu}\leftrightarrow \nu^s_{\mu}$, respectively.
In spite of such observations of the large mixing 
$\sin^2 2\theta\simeq 1$ in the disappearance experiments, 
effective mixing parameters 
$\sin^2 2\theta^{\alpha\beta}$ 
in appearance experiments $\nu_\alpha\rightarrow\nu_\beta$
($\alpha, \beta = e, \mu, \tau$) are  highly suppressed.
\end{abstract}


\narrowtext

It has, for many years, been a longed-for goal in 
the particle physics to give a unified description
of masses and mixings of quarks and leptons.
The most crucial clue to the unified description is 
in the investigation of the neutrino mass matrix.
Recent remarkable progress in the experimental studies of
the solar, atmospheric and terrestrial neutrinos has put the
realistic investigation of the neutrino mass matrix model
within our reach.

As a promising model which gives a unified 
description of quark and lepton mass matrices, the 
so-called ``universal seesaw" mass matrix model\cite{USMM} has 
recently revived.
The model has hypothetical fermions $F_i=U_i$, $D_i$, $N_i$
and $E_i$ ($i=1,2,3$) in addition to the conventional 
quarks and leptons $f_i=u_i$, $d_i$, $\nu_i$, $e_i$, where
these fermions belong to $f_L = (2,1)$, $f_R = (1,2)$, 
$F_L = (1,1)$ and $F_R = (1,1)$ of 
SU(2)$_L \times $SU(2)$_R$.
Note that the neutral leptons $N_i$ are the so-called 
``sterile" fermions because they are singlets of 
SU(2)$_L \times $SU(2)$_R$ and do not have U(1)-charges.
It has recently  been pointed out\cite{Knu98} by one of the authors 
(YK)  that such a model can yield three massless sterile 
neutrinos $\nu^s_{i}\equiv (N_{Li}-N_{Ri}^c)/\sqrt{2}$ 
[$N_R^c\equiv (N_R)^c\equiv C\overline{N}_R^T$] in 
a limit, and the sterile neutrinos $\nu^s_{i}$ compose
pseudo-Dirac neutrinos\cite{psD} $\nu_{i\pm}^{ps} \simeq
(\nu_{i}\pm \nu^s_{i})/\sqrt{2}$ with the masses $m(\nu_{i+}^{ps})
\simeq m(\nu_{i-}^{ps})$ together with the conventional
left-handed neutrinos $\nu_i\equiv\nu_{Li}$.

The purposes of the present paper are  to investigate 
an explicit model with three sterile neutrinos within 
the framework of the universal seesaw mass matrix model 
and to  give numerical predictions for neutrino masses
and mixings.
The solar\cite{solar} and atmospheric\cite{atm} data will be explained
by the mixings $\nu_{e}\leftrightarrow\nu^s_{e}$ and 
$\nu_{\mu}\leftrightarrow\nu^s_{\mu}$, respectively.
A similar idea has recently been proposed by Geiser\cite{Geiser}, 
who has introduced three sterile neutrinos 
phenomenologically.
However, in the present model, our sterile neutrinos 
$\nu^s_{i}$ come from the sterile fermions $N_i$ in the
universal seesaw mass matrix model, 
so that their masses and mixings are strongly constrained
by the parameters determined at the charged lepton masses 
and quark masses.

The seesaw mechanism  was first proposed\cite{nuSeesw} in order to 
answer the question of why neutrino masses are so invisibly small.
Then, in order to understand that the observed quark and lepton 
masses are considerably smaller than the electroweak scale 
$\Lambda_L$,
the mechanism was applied to the quarks\cite{USMM}:
A would-be seesaw mass matrix for $(f, F)$ is
expressed as
\begin{equation}
M = \left(\begin{array}{cc}
0 & m_L \\
m_R & M_F \\ 
\end{array} \right) = m_0 \left( 
\begin{array}{cc}
0 & Z_L \\
\kappa Z_R & \lambda Y_f \\
\end{array} \right) \ \ , 
\end{equation}
where the matrices $Z_L$, $Z_R$ and $Y_f$ are of the order one.
The matrices $m_L$ and $m_R$ take universal structures for
quarks and leptons.
Only the heavy fermion matrix $M_F$ takes a structure 
dependent on $f=u,d,\nu,e$.
For the case $\lambda \gg \kappa \gg 1$, 
the mass matrix (1) leads to the well-known seesaw expression
of the $3\times 3$ mass matrix for the fermions $f$:
\begin{equation}
 M_f \simeq -m_L M_F^{-1} m_R  \ .
\end{equation}
However, the observation of the top quark of 1994\cite{CDF} 
once aroused a doubt that the observed fact
$m_t \simeq 180$ GeV $\sim \Lambda_L = O(m_L)$ was not able to 
be understood from the universal seesaw mass matrix 
scenario, because $m_t\sim O(m_L)$ means $M_F^{-1}m_R \sim O(1)$.
For this question, a recent study\cite{KFzp,Morozumi} has given the answer that
the universal seesaw scenario is still applicable to top quark
if we put an additional constraint 
\begin{equation}
{\rm det}M_F =0 \ ,
\end{equation}
on the up-quark sector ($F=U$).
The light quarks $(q_{Li},q_{Ri})$ acquire the well-known 
seesaw masses of the order of $\Lambda_L \Lambda_R/\Lambda_S$
though the heavy fermion masses $m(Q_{Li},Q_{Ri})\sim \Lambda_S$
[$\Lambda_L\equiv m_0 =O(m_L)$, 
$\Lambda_R\equiv \kappa m_0 =O(m_R)$, and
$\Lambda_S\equiv \lambda m_0 =O(M_F)$], 
while the third up-quark $u_{L3}$ acquires a mass of the order of
$\Lambda_L$ together with the partner $U_{R3}$ 
since the seesaw mechanism does not work for 
the fermions $(u_{L3}, u_{R3})$ because of $m(U_{L3},U_{R3})= 0$.
Thus, by regarding the fermion state $(u_{L3}, U_{R3})$ 
[not $(u_{L3},u_{R3})$] as the top quark, we will easily be able 
to understand why only top quark $t$ 
acquires  the mass $m_t \sim O(m_L)$.
A suitable choice\cite{KFzp} of the matrix forms of $Z_L$, $Z_R$ and $Y_f$
can give reasonable values of the quark masses and 
Cabibbo-Kobayashi-Maskawa\cite{CKM} (CKM)  matrix parameters 
in terms of charged lepton masses. 

On the other hand, in the conventional universal seesaw mass matrix 
model\cite{USMM},
the light neutrino mass matrix $M_\nu$ is given by\cite{m-nu}
\begin{equation}
M_\nu \simeq - m_L M_N^{-1} m_L^T \ , 
\end{equation}
so that the smallness of the neutrino masses $m_\nu$ is understood
by a large value of $\kappa\equiv \Lambda_R/\Lambda_L$ 
($\kappa \gg 1$) because of $m_\nu \sim \Lambda_L^2/\Lambda_S
\sim (1/\kappa) m_{charged\ lepton}$.
The case (4) can be brought, for example, by assuming 
that the neutral leptons $N_{Li}$ and
$N_{Ri}$ which are singlets of SU(2)$_L\times$SU(2)$_R$
and have no U(1) charges acquire Majorana masses $M_M 
\sim \Lambda_S$ in addition to the Dirac masses 
$M_D\equiv M_N\sim \Lambda_S$ and assuming 
$M_M=M_D$ \cite{Koide-mnu}:
\begin{equation}
M_\nu=\left(
\begin{array}{cccc}
0 & 0 & 0 & m_L \\
0 & 0 & m_R^T & 0 \\
0 & m_R & M_M & M_D \\
m_L^T & 0 & M_D^T & M_M 
\end{array} \right) \ , 
\end{equation}
where $M_\nu$ is the mass matrix which is sandwiched between 
$(\overline{\nu}_L, \overline{\nu}_R^c, 
\overline{N}_L, \overline{N}_R^c)$ 
and $(\nu_L^c, \nu_R, N_L^c, N_R)^T$, and  
$\nu_L^c \equiv (\nu_L)^c \equiv C \overline{\nu}_L^T$ 
and so on.
The model given in Ref.~\cite{Koide-mnu} can readily lead to a large
mixing between $\nu_\mu$ and $\nu_\tau$ 
[or $\nu_e$ and $\nu_\mu$] with $m_1^\nu\ll m_2^\nu \simeq
m_3^\nu$ [or $m_1^\nu\simeq m_2^\nu \ll m_3^\nu$],
so that the model is favorable to the explanation of the
atmospheric neutrino data\cite{atm} which have suggested a 
large neutrino mixing $\nu_\mu \leftrightarrow \nu_x$.
However, the model is hard to give a simultaneous 
explanation of the atmospheric\cite{atm} and solar\cite{solar}
neutrino data.

We want to give a simultaneous 
explanation of the atmospheric and solar neutrino data.
The essential idea is as follows\cite{Knu98}:
In the present model, there is no quantum number which
distinguishes $N_L$ from $N_R$, because both fields are 
SU(2)$_L\times$SU(2)$_R$ singlets and do not have 
U(1) charges.
Therefore, if $\nu_L$ ($\nu_R$) acquire masses $m_L$ ($m_R$)
together with the partners $N_R$ ($N_L$), they may also 
acquire masses $m'_L$ ($m'_R$) together with the partners 
$N_L^c$ ($N_R^c$).
Then, the mass matrix for the neutrino sector is given by
\begin{equation}
M=\left(
\begin{array}{cccc}
0 & 0 & m'_L & m_L \\
0 & 0 & m_R^T & m^{\prime T}_R \\
m^{\prime T}_L & m_R & M_M & M_D \\
m_L^T & m'_R & M_D^T & M_M 
\end{array} \right) \ . 
\end{equation}
By rotating the fields 
$(\nu_L^c, \nu_R, N_L^c, N_R)$ by 
\begin{equation}
R_{34}=\left(
\begin{array}{cccc}
1 & 0 & 0 & 0 \\
0 & 1 & 0 & 0 \\
0 & 0 & \frac{1}{\sqrt{2}} & -\frac{1}{\sqrt{2}} \\
0 & 0 & \frac{1}{\sqrt{2}} & \frac{1}{\sqrt{2}} 
\end{array}\right) \ , 
\end{equation}
the $12\times 12$ mass matrix (6) is expressed as 
\begin{equation}
M'= R_{34} M R_{34}^T =  \left(
\begin{array}{cccc}
0 & 0 & \sqrt{2}\Delta_L & \sqrt{2}\overline{m}_L \\
0 & 0 & \sqrt{2}\Delta_R^T & \sqrt{2}\overline{m}_R^T \\
\sqrt{2}\Delta_L^T & \sqrt{2}\Delta_R & M_M-M_D & 0 \\
\sqrt{2}\overline{m}_L^T & \sqrt{2}\overline{m}_R 
& 0 & M_M+M_D 
\end{array} \right) \ ,
\end{equation}
where we have used $M_D^T=M_D$, and 
\begin{equation}
2 \Delta_L = m'_L - m_L \ , \ \ \ 
2 \Delta_R = m_R - m'_R \ , 
\end{equation}
\begin{equation}
2 \overline{m}_L = m'_L + m_L \ , \ \ \ 
2 \overline{m}_R = m'_R + m_R \ .
\end{equation}

In the $N_L\leftrightarrow N_R^c$ symmetric limit, i.e., 
in the limit of  $m'_L=m_L$,  $m'_R=m_R$ and 
$M_M=M_D\equiv M_N$, 
the mass matrices for the neutrinos $\nu^s$, $N^s$ and
$\nu_R$ are given by
\begin{equation}
M(\nu^s)=0 \ ,
\end{equation}
\begin{equation}
M(N^s)\simeq 2 M_N \ ,
\end{equation}
\begin{equation}
M(\nu_R) \simeq - m_R^T M_N^{-1} m_R \ ,
\end{equation}
where $\nu^s$ and $N^s$ are defined by
\begin{equation}
\nu^s_i=\frac{1}{\sqrt{2}}(N_{Li} - N_{Ri}^c) \ , 
\ \ \ 
N^s_{i}=\frac{1}{\sqrt{2}}(N_{Li} + N_{Ri}^c) \ .
\end{equation}
Furthermore, for a model with the relation $m_R\propto m_L$
such as a model given in Ref.~\cite{KFzp}, we obtain
\begin{equation}
M(\nu_L) =0 \ .
\end{equation}
Only when we assume a sizable difference between 
$m'_L$ and $m_L$ (and also between 
$m'_R$ and $m_R$), we can obtain visible neutrino
masses $m(\nu_L)\neq 0$ and $m(\nu^s)\neq 0$.

For $\Delta_L\neq 0$ and $\Delta_R\neq 0$, 
by using a unitary matrix $U_{12}$
\begin{equation}
U_{12}=\left(
\begin{array}{cccc}
C & -S & 0 & 0 \\
S^\dagger & C^\dagger & 0 & 0 \\
0 & 0 & 1 & 0 \\
0 & 0 & 0 & 1 
\end{array}\right) \ , 
\end{equation}
[$CC^\dagger + SS^\dagger =1$], 
we can express the mass matrix (8) as 
\begin{equation}
M^{\prime\prime}= U_{12} M' U_{12}^T =  \left(
\begin{array}{cccc}
0 & 0 & A & 0 \\
0 & 0 & B & G \\
A^T & B^T & 0 & 0 \\
0 & G^T & 0 & 2 M_N 
\end{array} \right) \ ,
\end{equation}
where we have put $M_M=M_D\equiv M_N$ and
\begin{equation}
A=\sqrt{2}(C \Delta_L -S \Delta_R^T) \ ,
\end{equation}
\begin{equation}
B=\sqrt{2}(S^\dagger \Delta_L +C^\dagger \Delta_R^T) \ ,
\end{equation}
\begin{equation}
G=\sqrt{2}(S^\dagger \overline{m}_L +C^\dagger 
\overline{m}_R^T) \ ,
\end{equation}
\begin{equation}
C S^{-1}= \overline{m}_R^T (\overline{m}_L)^{-1} \ .
\end{equation}
Since $M_N \sim \Lambda_S$, $G\sim \Lambda_R$, 
$O(B)\ll \Lambda_R$ and $O(A)\ll \Lambda_L$, 
by using the seesaw approximation,
we obtain the $9\times 9$ mass matrix for
states $(\nu_L, \nu_R, \nu^s)$ 
\begin{equation}
M^{(9)}\simeq \left(
\begin{array}{ccc}
0 & 0 & A \\
0 & -G(2M_N)^{-1} G^T & B \\
A^T & B^T & 0 
\end{array} \right) \ ,
\end{equation}
together with the $3\times 3$ mass matrix (12)
for the neutral leptons $N^s$.
Furthermore, for $O(GM_N^{-1}G^T)=\Lambda_R^2/\Lambda_S
=(\kappa^2/\lambda)m_0 \gg O(A),O(B)$,
we obtain the $6\times 6$ mass matrix for 
$(\nu_L, \nu^s)$ 
\begin{equation}
M^{(6)}\simeq \left(
\begin{array}{cc}
0 &  A \\
A^T & 2B(G^T)^{-1}M_N G^{-1}B^T  \\
\end{array} \right) \ ,
\end{equation}
together with the $3\times 3$ mass matrix
for the right-handed neutrinos $\nu_R$
\begin{equation}
M(\nu_R)\simeq -G(2M_N)^{-1} G^T \ .
\end{equation}

If we choose $O(A)\gg O(B(G^T)^{-1}M_N G^{-1}B^T)$,
we obtain three sets of the pseudo-Dirac neutrinos
$(\nu^{ps}_{i+}, \nu^{ps}_{i-})$ ($i=e, \mu, \tau$)
which are large mixing states between $\nu_{i}$ 
and $\nu^s_{i}$, i.e.,
\begin{equation}
\nu^{ps}_{i\pm} \simeq \frac{1}{\sqrt{2}}
(\nu_{i}\pm \nu^s_i) \ ,
\end{equation}
and whose masses are almost degenerate.

So far, we have not assumed explicit structures of the 
matrices $Z_L$, $Z_R$ and $Y_f$.
Here, in order to give a realistic numerical example,
we adopt a model with  special forms of $Z_L$, $Z_R$ and $Y_f$
which can lead to successful quark masses and mixings.
The model is based on the following working hypotheses\cite{KFzp}: 

\noindent
(i) The matrices $Z_L$ and $Z_R$, which are universal
for quarks and leptons, have the same 
structure:
\begin{equation}
Z_L = Z_R \equiv Z = {\rm diag} (z_1, z_2, z_3) \ \ , 
\end{equation}
with $z_1^2 + z_2^2 + z_3^2 = 1$, 
where, for convenience, we have taken a basis on which 
the matrix $Z$ is diagonal. 

\noindent
(ii) The matrices $Y_f$, which have structures 
dependent on the fermion sectors $f=u,d,\nu,e$, take
a simple form [(unit matrix)+(a rank one matrix)]:
\begin{equation}
Y_f = {\bf 1} + 3 b_f X \ \ . 
\end{equation}
(iii) The rank one matrix $X$ is  given by
a democratic form
\begin{equation}
X = \frac{1}{3}\left(\begin{array}{ccc}
1 & 1 & 1 \\
1 & 1 & 1 \\
1 & 1 & 1 \\
\end{array} \right) \  , 
\end{equation}
on the family-basis where the matrix $Z$ is diagonal.

\noindent
(iv) In order to fix the parameters $z_i$, we 
tentatively take $b_e = 0$ for the charged lepton sector,
so that the parameters $z_i$ are given by
\begin{equation}
\frac{z_1}{\sqrt{m_e}} = \frac{z_2}{\sqrt{m_\mu}} = 
\frac{z_3}{\sqrt{m_\tau}} = \frac{1}{\sqrt{m_e + m_\mu + m_\tau}} \  , 
\end{equation}
from $M_e \simeq m_L M_E^{-1} m_R =(\kappa/\lambda) m_0 Z\cdot 
{\bf 1} \cdot Z$.

The mass spectra are essentially characterized by the parameter $b_f$.
We take $b_u = -1/3$  for up-quark sector, 
because, at $b_u=-1/3$, we can obtain the maximal top-quark mass 
enhancement $m_t \simeq ({1}/{\sqrt{3}}) m_0$,
and  a successful relation 
${m_u}/{m_c} \simeq ({3}/{4})({m_e}/{m_\mu})$
independently of the value of $\kappa/\lambda$. 
The value of $\kappa/\lambda$ is determined from the observed ratio
$m_c/m_t$ as $\kappa/\lambda=0.0198$. 
Considering the successful relation 
${m_d m_s}/{m_b^2}\simeq 4 {m_e m_\mu}/{m_\tau^2}$ 
for $b_d\simeq -1$, we seek for the best fit point of
$ b_d=- e^{i\beta_d}$ ($\beta_d^2 \ll 1$).
The observed ratio $m_d/m_s$ fixes the value $\beta_d$ 
as $\beta_d=18^\circ$.  
Then we can obtain\cite{KFzp} the reasonable quark mass ratios,
not only $m_i^u/m_j^u$, $m_i^d/m_j^d$,  but also 
$m_i^u/m_j^d$:
$m_u=0.00234$ GeV, $m_c=0.610$ GeV, 
$m_t=0.181$ GeV, $m_d=0.00475$ GeV, $m_s=0.0923$ GeV, 
$m_b=3.01$ GeV, where we have taken 
${(m_0\kappa/\lambda)_q}/{(m_0\kappa/\lambda)_e}=3.02$
in order to fit the observed quark mass values at 
$\mu=m_Z$\cite{q-mass}.
We also obtain the reasonable values of the CKM
matrix parameters: 
$|V_{us}|=0.220$, $|V_{cb}|=0.0598$,  
$|V_{ub}|=0.00330$, $|V_{td}|=0.0155$. 
(The value of $|V_{cb}|$ is somewhat larger than the observed value.
For the improvement of the numerical value, see Ref.~\cite{KFptp}.)

Stimulated by this phenomenological success in the present 
model for the quark masses and mixings, we apply the model with
(26) - (28) to the neutrino mass matrix (6).
For simplicity, we consider the case
\begin{equation}
m'_L=(1-\varepsilon_L) m_L \ , \ \ \ 
m'_R=(1-\varepsilon_R) m_R \ .
\end{equation}
Then, we obtain
\begin{equation}
A\simeq \frac{1}{\sqrt{2}}m_0 (\varepsilon_R+
\varepsilon_L-\varepsilon_L\varepsilon_R) Z \ ,
\end{equation}
\begin{equation}
B\simeq \frac{1}{\sqrt{2}}m_0 \kappa \left(
\varepsilon_R-\frac{1}{\kappa^2}\varepsilon_L \right) Z 
\ ,
\end{equation}
\begin{equation}
G\simeq \sqrt{2} m_0\kappa\left(1-\frac{1}{2}\varepsilon_R
+\frac{1}{\kappa^2}\right) Z \ ,
\end{equation}
\begin{equation}
M_N =m_0 \lambda ({\bf 1}+3b_\nu X) \ .
\end{equation}
What is of great interest to us is to evaluate 
the masses of the pseudo-Dirac neutrino states,
whose mass matrix is given by (23):
\begin{equation}
M^{(6)}\simeq \frac{1}{\sqrt{2}}m_0 
(\varepsilon_R+\varepsilon_L) \left(
\begin{array}{cc}
0 & Z \\
Z & \rho Y_\nu 
\end{array}\right) \ ,
\end{equation}
where
\begin{equation}
\rho \simeq\frac{1}{\sqrt{2}} \frac{\varepsilon_R^2}{
\varepsilon_R+\varepsilon_L}\lambda \ .
\end{equation}
For $Z\gg |\rho Y_\nu|$, i.e., $z_i\gg |\rho (1+b_\nu)|$,
we obtain the masses $m(\nu_{i\pm}^{ps})$ 
of the pseudo-Dirac neutrinos $\nu_{i\pm}^{ps}$
\begin{equation}
m(\nu_{i\pm}^{ps})\simeq \frac{1}{\sqrt{2}}m_0 
(\varepsilon_R+\varepsilon_L)\left( z_i
\pm \rho \frac{1+b_\nu}{2}\right) \ .
\end{equation}
Here, the double sign, $\pm$, in the right-hand side
of (37) does not always correspond to the double sign
of $\nu_{i\pm}^{ps}$ in the left-hand side.
For the mass eigenstates $\nu_{i\pm}^{ps}$ defined by (25),
the order of the magnitudes of the mass eigenvalues are
$m(\nu_{e-}^{ps})<m(\nu_{e+}^{ps})<m(\nu_{\mu +}^{ps})
<m(\nu_{\mu -}^{ps})<m(\nu_{\tau -}^{ps})<m(\nu_{\tau +}^{ps})$
(see Fig.~1).
On the other hand, the squared mass differences 
$\Delta m^2(\nu_{i}^{ps})=|m^2(\nu_{i+}^{ps})
-m^2(\nu_{i-}^{ps})|$ are given by
\begin{equation}
\Delta m^2(\nu_{i}^{ps}) \simeq 
m_0^2 (\varepsilon_R+\varepsilon_L)^2
|\rho|z_i
 \left[( 1+b_\nu) +\rho^2 b_\nu^2
\frac{2 b_\nu z_i^2+(1+b_\nu)(2z_i^2 -z_j^2-z_k^2)}{
(z_i^2-z_j^2)(z_i^2-z_k^2)} \right] \ ,
\end{equation}
where $(i,j,k)$ is a cyclic permutation of $(i,j,k)=(1,2,3)$.

As the numerical input, we use 
\begin{equation}
\left(m_0\frac{\kappa}{\lambda}\right)_\ell\equiv 
\left(m_0\frac{\kappa}{\lambda}\right)_e={\rm Tr}M_e
=(m_\tau+m_\mu+m_e)_{\mu=m_Z}=1.8499 \ {\rm GeV} \ ,
\end{equation}
\begin{equation}
(\kappa/\lambda)_\ell \equiv (\kappa/\lambda)_q=0.02 \ ,
\end{equation}
so that
\begin{equation}
m_0=0.925\times 10^{11} \ {\rm eV} \ .
\end{equation} 

For simplicity, we seek for the solutions for the 
case $\varepsilon_L =\varepsilon_R\equiv \varepsilon$, 
so that the parameters in the neutrino masses and 
mixings are three, i.e., $\varepsilon$, $b_\nu$ and
$\lambda$.
We tentatively choose $m(\nu_{\tau\pm}^{ps})\simeq 5$ 
eV, which is suggested from a cosmological model with
cold+hot dark matter (CHDM)\cite{CHDM}.
Then, for $b_\nu\simeq -1$, the parameter value 
$\varepsilon$ is given from (37) as
\begin{equation}
\varepsilon \equiv \varepsilon_L=\varepsilon_R\simeq 3.8 
\times 10^{-11} \ .
\end{equation}
The recent atmospheric data from Super-Kamiokande\cite{Kajita}
have suggested the neutrino oscillation 
$\nu_\mu\rightarrow\nu_\tau$ with 
$\Delta m^2 \simeq 2.2 \times 10^{-3}$ eV$^2$ and
$\sin^2 2\theta\simeq 1$.
Since the value of $\Delta m^2$ in the 
case of $\nu_\mu \rightarrow \nu^s_\mu$ is given by
$\Delta m^2(\nu_\mu\rightarrow\nu^s_\mu)\simeq 
2\times \Delta m^2(\nu_\mu\rightarrow\nu_\tau)$\cite{Liu},
we put $\Delta m^2(\nu_\mu^{ps})=\Delta m^2_{atm}
\simeq 5\times 10^{-3}$ eV$^2$.
Then, from (38), for $b_\nu\simeq -1$, we obtain
\begin{equation}
\rho\simeq 0.057\ , \ \ \ \lambda\simeq 4.3\times 10^{9} \ 
\ (\kappa\simeq 8.5\times 10^{7}) \ .
\end{equation}
The numerical results of the masses 
$m(\nu_{i}^{ps})\equiv 
(m(\nu_{i+}^{ps})+m(\nu_{i-}^{ps}))/2$ 
for the case of $b_\nu\simeq -1$ are as follows:
\begin{equation}
m(\nu_{e}^{ps}) \simeq 0.078\ {\rm eV}\ , \ \ \ 
m(\nu_{\mu}^{ps}) \simeq 1.2\ {\rm eV}\ , \ \ \ 
m(\nu_{\tau}^{ps})\simeq 4.8\ {\rm eV}\ .
\end{equation}

The value of $\Delta m^2(\nu_e^{ps})$ is highly 
sensitive to the parameter $b_\nu$.
The large-angle Mikheyev-Smirnov-Wolfenstein (MSW) 
solutions\cite{MSW} of solar neutrino problem, 
$(\Delta m^2,\sin^2 2\theta)\simeq (10^{-5}\ {\rm eV}^2, 0.6)$,
is ruled out for the case $\nu_e\rightarrow\nu^s_e$\cite{Hata}.
Therefore,  we use
the vacuum solution $(\Delta m^2, \sin^2 2\theta)
\simeq (0.6-0.8\times 10^{-10}\ {\rm eV}^2, 1)$\cite{vacuum,Faid}
in order to fix the parameter $b_\nu$. 
For example, the value $b_\nu+1=3.078\times 10^{-5}$ gives 
$(\Delta m^2, \sin^2 2\theta)\simeq (0.605\times 10^{-5}\ {\rm eV}^2,
0.9997)$, which is safely in the allowed region obtained by Faid 
{\it et al.}\cite{Faid} by using the solar neutrino data from 
Super-Kamiokande and Borexino.
In Table I, 
the neutrino oscillation parameters $\Delta m^2_{ij}$
and $\sin^2 2\theta_{ij}^{\alpha\beta}$ are listed.
Here, the neutrino oscillation $P(\nu_\alpha\rightarrow
\nu_\beta)$ is given by
\begin{equation}
P(\nu_\alpha\rightarrow\nu_\beta(\nu^s_\beta))\simeq 
\sum_{i=e,\mu,\tau} \sin^2 2\theta^{\alpha\beta}_{ii}
\sin^2 \frac{L\Delta m^2(\nu_i^{ps})}{4E} 
+\sum_{i>j} \sin^2 2\theta^{\alpha\beta}_{ij}
\sin^2 \frac{L\Delta m^2_{ij}}{4E} \ ,
\end{equation}
where $\Delta m^2(\nu_i^{ps})=|m^2(\nu_{i+}^{ps})
-m^2(\nu_{i-}^{ps})|$, 
$\Delta m^2_{ij}=m^2(\nu_{i}^{ps})-m^2(\nu_j^{ps})$, 
\begin{equation}
\sin^2 2\theta^{\alpha\beta}_{ii}=
-4 {\rm Re}\left(U_{\alpha, i+}U^*_{\alpha, i-}
U^*_{\beta, i+}U_{\beta, i-}\right) \ ,
\end{equation}
\begin{equation}
\sin^2 2\theta^{\alpha\beta}_{ij}
=-4 {\rm Re}\left[(U_{\alpha, i+}U^*_{\beta, i+}+
U^*_{\alpha, i-}U_{\beta, i-})
(U_{\alpha, j+}U^*_{\beta, j+}+
U^*_{\alpha, j-}U_{\beta, j-})\right] \ ,
\end{equation}
and we have used the approximation 
$ m^2(\nu_{i+}^{ps})- m^2(\nu_{j+}^{ps}) \simeq 
m^2(\nu_{i+}^{ps})- m^2(\nu_{j-}^{ps}) \simeq 
m^2(\nu_{i-}^{ps})- m^2(\nu_{j+}^{ps}) \simeq 
m^2(\nu_{i-}^{ps})- m^2(\nu_{j-}^{ps}) \equiv
\Delta m^2_{i j} $ ($i,j=e,\mu,\tau$).
Also, since $\Delta m^2_{\tau\mu}\gg \Delta m^2_{\mu e}$,
i.e., $\Delta m^2_{\tau\mu}\simeq \Delta m^2_{\tau e}$,
in Table I, 
we have denoted the sum 
$\sin^2 2\theta^{\alpha\beta}_{\tau\mu}+
\sin^2 2\theta^{\alpha\beta}_{\tau e}$ instead of 
$\sin^2 2\theta^{\alpha\beta}_{\tau\mu}$ and 
$\sin^2 2\theta^{\alpha\beta}_{\tau e}$ 
as the effective mixing parameter $\sin^2 2\theta$
for $\Delta m^2_{\tau\mu}$. 

As seen in Table I, at $\Delta m^2\simeq 0.605\times 10^{-10}$
eV$^2$, the oscillation $\nu_e\rightarrow \nu_x$ is
caused not only by $\nu_e\rightarrow \nu_e^s$ 
($\sin^2 2\theta^{e s}_{e e}=0.941$)
but also by $\nu_e\rightarrow \nu_\mu$
($\sin^2 2\theta^{e \mu}_{e e}=0.055$).
The small parameter $b_\nu+1$ is insensitive to the 
numerical results except for 
$\Delta m^2(\nu_e^{ps})$.
For example, by taking $b_\nu+1=2.8\times 10^{-4}$,
we obtain $\Delta m^2(\nu_e^{ps})=1.0\times 10^{-5}$
eV$^2$, while the other numerical results, 
$\Delta m^2$ and $\sin^2 2\theta$, are almost 
unchanged.
The value $\sin^2 2\theta^{e \mu}_{e e}=0.055$
at $\Delta m^2\simeq 10^{-5}$ eV$^2$ is too
large compared with the small angle solution
$\sin^2 2\theta\simeq 7\times 10^{-3}$ in the MSW
solutions\cite{MSW,Hata}, we cannot accommodate 
the present model to the MSW solutions.

Thus, the model can explain the solar and atmospheric 
neutrino data by the large mixings $\nu_{e}\leftrightarrow
\nu^s_{e}$ and $\nu_{\mu}\leftrightarrow \nu^s_{\mu }$, 
respectively.
On the other hand, in the appearance experiments
$\nu_\alpha\rightarrow\nu_\beta$ ($\alpha, \beta=e,\mu,\tau$),
the effective mixing parameters 
$\sin^2 2\theta_{ij}^{\alpha\beta}$ defined by (47)
are highly suppressed 
because of  the cancellation between $\nu_{i +}^{ps}$
and $\nu_{i -}^{ps}$. 
Therefore, in spite of a suitable value of $\Delta m^2$, 
$\Delta m^2_{e\mu}\simeq 1.4$ eV$^2$,
the present model cannot explain the neutrino oscillation
$\overline{\nu}_\mu\rightarrow\overline{\nu}_e$ 
experiment by the liquid scintillator neutrino detector 
(LSND)\cite{LSND} at Los Alamos because our prediction is
$\sin^2 2\theta^{e\mu}_{e\mu}=1.7\times 10^{-7}$, 
although Geiser has explained the LSND data by the
oscillation $\nu_e\leftrightarrow\nu_\mu$ in his 
model with three sterile neutrinos\cite{Geiser}.
The effective mixing angles
$\sin^2 2\theta^{\mu\tau}= 3\times 10^{-6}$ at
$\Delta m^2\simeq 22$ eV$^2$ and
$\sin^2 2\theta^{\mu\tau}= 0.0008$ at
$\Delta m^2\simeq 1.4$ eV$^2$ are too
small to be detected by 
CHORUS\cite{CHORUS} and NOMAD\cite{NOMAD} experiments at CERN. 
The effective electron neutrino mass
$\langle m_\nu\rangle$ for the case in Table I is also 
suppressed:
\begin{equation}
\langle m_\nu\rangle =\left|\sum_{i=1}^6 m_i^\nu U_{ei}^2
\right| = 2.8\times 10^{-6} \ {\rm eV} \ ,
\end{equation}
which is safely small compared with the upper bound
on $\langle m_\nu\rangle$ from the neutrino double
$\beta$ decays $\langle m_\nu\rangle < 0.68$ eV\cite{bb}.

The numerical results in Table I should not be 
taken rigidly. 
These values are sensitive to the parameter values used here,
especially the value $\Delta m^2(\nu_{e}^{ps})$ is to the 
parameter $(b_\nu +1)$.
The values are only an example.
We can give any values as far as these values are within the
similar orders.
Furthermore, we can also obtain similar results by using
another parametrization
\begin{equation}
m'_L= m_L e^{i\varepsilon_L}\ , \ \ \ 
m'_R= m_R e^{i\varepsilon_R}\ ,
\end{equation}
instead of (30) (the numerical values of $\varepsilon_L$ 
and  $\varepsilon_R$ are different from those in the 
parametrization (30)).

Generally, a model with sterile neutrinos is stringently
constrained by the big bang nucleosynthesis (BBN) 
scenario.
However, it has recently been pointed out by Foot and
Volkas\cite{Foot} that the stringent bounds can 
considerably be reduced by the lepton number asymmetry
in the early Universe.
Our pseudo-Dirac neutrinos $\nu_{\tau\pm}^{ps}$ and
$\nu_{\mu\pm}^{ps}$ have masses greater than 1 eV, 
and the squared mass difference $\Delta m^2_{\tau\mu}$
is $\Delta m^2_{\tau\mu}\simeq 22$ eV$^2$.
The mixing angles except those for 
$\nu_i\leftrightarrow\nu_i^s$ are sufficiently small.
Therefore, the present model does not spoil the 
BBN scenario.

In conclusion, we have discussed  a possible 
neutrino-mass-generation scenario within the framework of the
universal seesaw mechanism, especially, on the basis
of a model\cite{KFzp} with  special matrix forms of $m_L$, $m_R$ and
$M_F$, which can answer the question why only top quark $t$ 
acquires the mass of the order of the electroweak scale 
$\Lambda_L=O(m_L)$, and can give reasonable quark 
masses and mixings in terms of the charged lepton 
masses.
The model provides three sets of the light pseudo-Dirac 
neutrinos $(\nu_{i+}^{ps},\nu_{i-}^{ps})$ ($i=e,\mu,\tau$)
are mixing states $\nu_{i\pm}^{ps}\simeq (\nu_i\pm \nu_i^s)
/\sqrt{2}$ between the conventional left-handed neutrinos 
$\nu_{i}$ and sterile neutrinos 
$\nu_i^s\equiv (N_{Li}-N_{Ri}^c)/\sqrt{2}$.
The mixings $\nu_{e}\leftrightarrow \nu_e^s$ and 
$\nu_{\mu}\leftrightarrow \nu_\mu^s$  
can give
reasonable explanations of the solar and atmospheric
neutrino data, respectively, with
$m(\nu_{\tau\pm}^{ps})\simeq 5$ eV.
Note that in spite of such a large mixing in 
a disappearance experiment, the  effective mixing 
parameters $\sin^2 2\theta^{\alpha\beta}_{ij}$ 
in the appearance experiments $\nu_\alpha\rightarrow\nu_\beta$
are highly suppressed.
This is the most remarkable feature in the model
with the pseudo-Dirac neutrinos.
If future appearance experiments $\nu_\mu\rightarrow\nu_\tau$
at MINOS\cite{MINOS}, ICARUS\cite{ICARUS} and K2K\cite{K2K}
rule out the region $(\Delta m^2, \sin^2 2\theta)\simeq 
(10^{-3}-10^{-2}\ {\rm eV}^2, 0.8-1)$ which is 
suggested by the atmospheric data, the present scenario
with the three sterile neutrinos $\nu_i^s$ will become
an promising candidate of the possible interpretation
of the discrepancy.

At present, except for the LSND data, 
there is no positive motivation
for such a model with three sterile neutrinos.
Since the recent KARMEN data\cite{KARMEN} seem to disfavor 
the LSND solution $(\Delta m^2, \sin^2 2\theta)
=(0.3\, {\rm eV}^2, 0.04)-(2\, {\rm eV}^2, 0.002)$,
the motivation of the model with the sterile neutrinos
seems to disappear.
However, the present model with three sterile neutrinos is motivated
not on the phenomenological interests in the neutrino data, but on
the phenomenological success\cite{KFzp,KFptp} of the universal seesaw mass matrix
model for quarks.
As far as the model is based on the SU(2)$_L\times$SU(2)$_R\times$U(1)$_Y$
symmetry, the model includes the sterile fermions $N_i$, so that we 
inevitably have three sterile neutrinos $\nu^s_i$ in the model.
If we do not want such the light sterile neutrinos $\nu^s_i$,
we must consider an additional mechanism which guarantees 
$m'_L=m'_R=0$.

The present scenario can bring fruitful phenomenology 
into the neutrino physics, but there are many adjustable parameters 
in the present model, and those parameters have been determined 
by hand and by way of trial.
In the present scenario, the active neutrinos $\nu_{i}$ and sterile 
neutrinos $\nu_i^s$ are massless in the limit of $N_L\leftrightarrow N_R^c$
symmetry, and the small neutrino masses are induced by the violation 
of the $N_L\leftrightarrow N_R^c$ symmetry, i.e., by the parameters 
$\varepsilon_L\neq 0$ and $\varepsilon_R\neq 0$. 
In other words, the neutrino
masses are originated in ``non-seesaw" mechanism in spite of the
scenario in the ``seesaw" model.
We consider that the deviation between $m'_L$ ($m'_R$) and 
$m_L$ ($m_R$) can come from an asymmetric evolution of the 
Yukawa coupling constants from a unification scale $\mu=\Lambda_X$
to the electroweak scale $\mu=\Lambda_L$.
However, whether this conjecture is reasonable or not is open 
question at present.

{\it Note added.} 
After completion of this manuscript the authors became aware of
a paper by Boweres and Volkas (hep-ph/9804310), in which 
a neutrino mass matrix model with three pseudo-Dirac neutrinos 
has also been proposed within the framework of the universal 
seesaw mass matrix model.
They have derived the light pseudo-Dirac neutrinos in their 
model with $m'_L=m'_R=0$.
Thier mass matrix resembles the model (5), but they assumed
that $O(M_M)\ll O(M_D)$.
Even if we accept the assumption  $O(M_M)\ll O(M_D)$, 
their idea is not applicable to our model, because their 
light pseudo-Dirac neutrinos have masses of the order of
$m_L M_D^{-1} m_R$ in our scheme, so that the pseudo-Dirac
neutrinos will acquire the masses of the same order as the 
charged leptons and light quarks have.
However, it is noticeable that the model with  $m'_L=m'_R=0$
can also lead to three pseudo-Dirac neutrinos.

\vspace*{.2in}

\acknowledgments

The authors would like to thank M.~Tanimoto for  
helpful discussions and his useful comments.
They also thank A.~Geiser, O.~Yasuda and D.~Suematsu
for useful comments on the big bang nucleosynthesis bounds
for a model with sterile neutrinos and informing 
Ref.\cite{Foot}, Q.~Y.~Liu for useful information
on the neutrino mass values in the cases 
$\nu_e\leftrightarrow\nu_e^s$ and 
$\nu_\mu\leftrightarrow\nu_\mu^s$,
T.~Hara and K.~Niwa for useful information on 
the CHORUS experiment, and
R.~R.~Volkas for informing his recent suggestive works.
This work was supported by the Grand-in-Aid for Scientific
Research, Ministry of Education, Science and Culture,
Japan (No.~08640386).

\vglue.2in

\newpage
\begin{table}[htb]
\caption{Neutrino oscillation parameters for the case:
$b_\nu+1=3.078\times 10^{-5}$, 
$\varepsilon_L=\varepsilon_R=3.79\times 10^{-11}$, 
and $\lambda=4.25\times 10^{9}$ ($\kappa=8.50\times 10^7$) .}

$$
\begin{array}{|c|c|c|c|c|c|}\hline\hline
  & \Delta m^2(\nu_e^{ps})= & \Delta m^2(\nu_\mu^{ps})= &
\Delta m^2(\nu_\tau^{ps})= & \Delta m^2_{\mu e}= &
\Delta m^2_{\tau\mu}= \\
  &  0.605\times 10^{-10} \ {\rm eV}^2 &
4.97\times 10^{-3} \ {\rm eV}^2 &
1.98\times 10^{-2} \ {\rm eV}^2 &
1.43 \ {\rm eV}^2 & 21.9 \ {\rm eV}^2 \\ \hline\hline
\nu_{\alpha}\rightarrow \nu_{\beta} & 
\sin^2 2\theta^{\alpha\beta}_{ee}= &
\sin^2 2\theta^{\alpha\beta}_{\mu\mu}= &
\sin^2 2\theta^{\alpha\beta}_{\tau\tau}= &
\sin^2 2\theta^{\alpha\beta}_{e\mu}= &
\begin{array}{c}
\sin^2 2\theta^{\alpha\beta}_{e\tau}+ \\
\sin^2 2\theta^{\alpha\beta}_{\mu\tau}=
\end{array} \\ \hline
\nu_{e}\rightarrow \nu^s_{e}  
& 0.941
& 1.4\times 10^{-5}
& 3.2\times 10^{-9}
& 6.1\times 10^{-8}
& -8.6\times 10^{-10} \\
\nu_{\mu}\rightarrow \nu^s_{\mu}  
& 1.4\times 10^{-5}
& 0.940
& 6.7\times 10^{-7}
& -3.4\times 10^{-9}
& 5.9\times 10^{-8} \\
\nu_{\tau}\rightarrow \nu^s_{\tau}  
& 2.7\times 10^{-9}
& 8.2\times 10^{-7}
& 0.992
& 3.1\times 10^{-9}
& 8.9\times 10^{-7} \\ \hline
\nu_{e}\rightarrow \nu_{\mu}  
& 0.055
& 0.00023
& -2.1\times 10^{-10}
& 1.7\times 10^{-7}
& 1.1\times 10^{-9}  \\
\nu_{\mu}\rightarrow \nu_{\tau}  
& -1.8\times 10^{-5} 
& 0.0034
& 0.00022
& 0.00076
& 3.0\times 10^{-6} \\
\nu_{e}\rightarrow \nu_{\tau}  
& 0.0032
& -9.0\times 10^{-7}
& 9.1\times 10^{-7}
& 3.6\times 10^{-6}
& 1.4\times 10^{-8}
\\ \hline
\nu_{e}\rightarrow \nu_{e}  
& 0.9997
& 6.2\times 10^{-8}
& 9.0\times 10^{-13}
& 0.00099
& 3.8\times 10^{-6} \\ 
\nu_{\mu}\rightarrow \nu_{\mu}  
& 0.0031
& 0.892
& 5.1\times 10^{-8}
& 0.210
& 0.00091 \\
\nu_{\tau}\rightarrow \nu_{\tau}  
& 1.0\times 10^{-5}
& 1.3\times 10^{-5}
& 0.986
& 5.3\times 10^{-5}
& 0.029  \\ \hline\hline
\end{array}
$$
\end{table}

\newpage

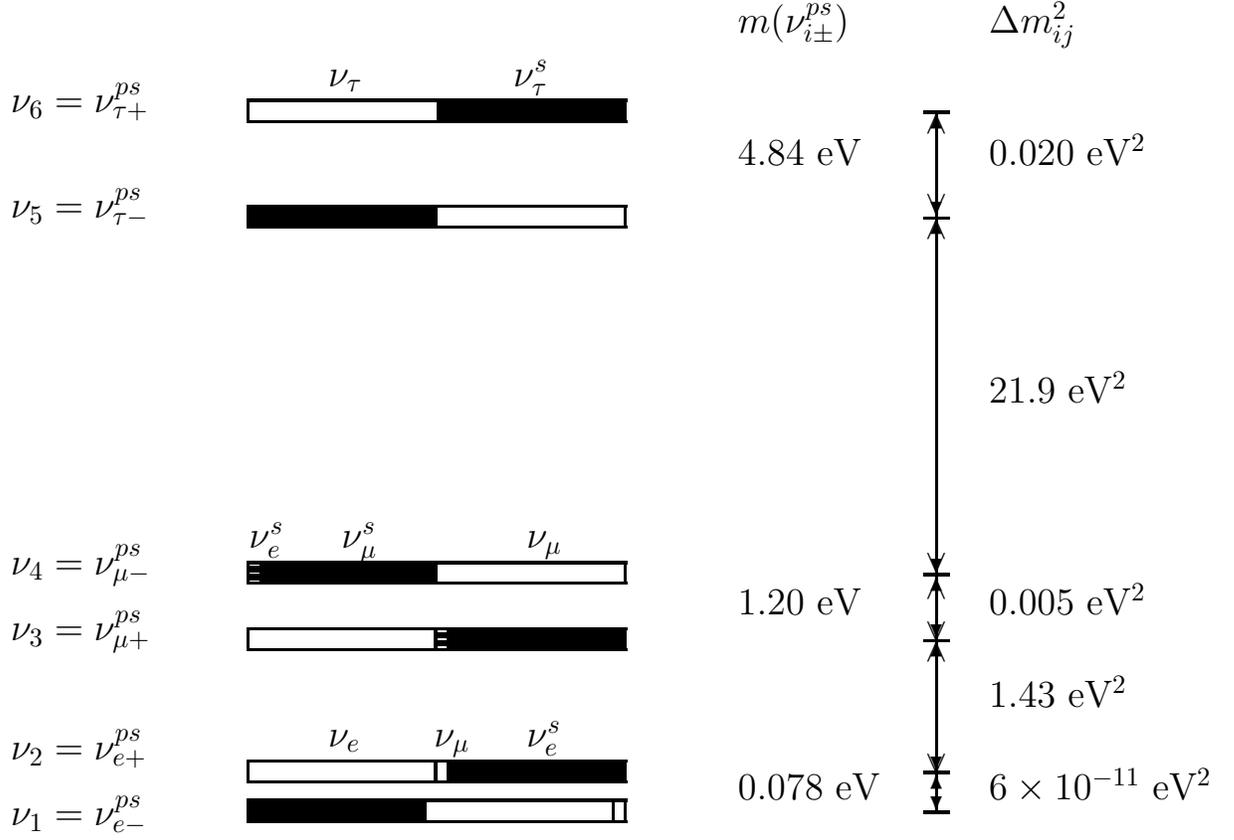
\begin{figure}
\caption{Mass spectrum of the pseudo-Dirac neutrinos
$\nu_{i\pm}^{ps}$.  The parameter values are the same as
those in Table I. Boxes correspond to different mass eigenstates.
The sizes of different regions in the boxes determine flavors of
mass eigenstates, $|U_{\alpha i}|^2$.
White and black regions correspond to the active and sterile flavors,
respectively.}

\begin{picture}(420,350)(0,0)
\setlength{\unitlength}{1pt}
\thicklines
\put(100,277){\framebox(142,7){}}
\put(171,277){\rule{71pt}{7pt}}
\put(100,237){\framebox(142,7){}}
\put(100,237){\rule{71pt}{7pt}}
\put(100,102){\framebox(142,7){}}
\put(100,102){\framebox(4,7){}}
\put(100,103){\rule{4pt}{2pt}}
\put(100,106){\rule{4pt}{2pt}}
\put(104,102){\rule{67pt}{7pt}}
\put(100,77){\framebox(142,7){}}
\put(171,77){\framebox(4,7){}}
\put(171,78){\rule{4pt}{2pt}}
\put(171,81){\rule{4pt}{2pt}}
\put(175,77){\rule{67pt}{7pt}}
\put(100,27){\framebox(142,7){}}
\put(171,27){\framebox(4,7){ }}
\put(175,27){\rule{67pt}{7pt}}
\put(100,12){\framebox(142,7){}}
\put(100,12){\rule{67pt}{7pt}}
\put(238,12){\framebox(4,7){ }}
%
\put(10,280){\large $\nu_{6}=\nu_{\tau +}^{ps}$}
\put(10,240){\large $\nu_{5}=\nu_{\tau -}^{ps}$}
\put(285,310){\large $m(\nu_{i \pm}^{ps})$}
\put(285,260){\large 4.84 eV}
\put(380,310){\large $\Delta m_{ij}^{2}$}
\put(380,260){\large 0.020 eV$^{2}$}

\put(380,170){\large 21.9 eV$^{2}$}

\put(10,105){\large $\nu_{4}=\nu_{\mu -}^{ps}$}
\put(10,80){\large $\nu_{3}=\nu_{\mu +}^{ps}$}
\put(285,90){\large 1.20 eV}
\put(380,90){\large 0.005 eV$^{2}$}

\put(380,55){\large 1.43 eV$^{2}$}

\put(10,35){\large $\nu_{2}=\nu_{e +}^{ps}$}
\put(10,10){\large $\nu_{1}=\nu_{e -}^{ps}$}
\put(285,20){\large 0.078 eV}
\put(380,20){\large $6\times 10^{-11}$ eV$^{2}$}
\put(355,280){\line(1,0){10}}
\put(355,240){\line(1,0){10}}
\put(355,105){\line(1,0){10}}
\put(355,80){\line(1,0){10}}
\put(355,30){\line(1,0){10}}
\put(355,15){\line(1,0){10}}
\put(360,260){\vector(0,1){20}}
\put(360,260){\vector(0,-1){20}}
\put(360,200){\vector(0,1){40}}
\put(360,200){\vector(0,-1){95}}
\put(360,95){\vector(0,1){10}}
\put(360,95){\vector(0,-1){15}}
\put(360,50){\vector(0,1){30}}
\put(360,50){\vector(0,-1){20}}
\put(360,20){\vector(0,1){10}}
\put(360,25){\vector(0,-1){10}}
\put(356,272){$\wedge$}
\put(356,242){$\vee$}
\put(356,232){$\wedge$}
\put(356,106){$\vee$}
\put(356,96){$\wedge$}
\put(356,80){$\vee$}
\put(356,72){$\wedge$}
\put(356,30){$\vee$}
\put(130,290){\large $\nu_\tau$}
\put(200,290){\large $\nu_\tau^s$}
\put(100,115){\large $\nu_e^s$}
\put(135,115){\large $\nu_\mu^s$}
\put(205,115){\large $\nu_\mu$}
\put(130,40){\large $\nu_e$}
\put(170,40){\large $\nu_\mu$}
\put(205,40){\large $\nu_e^s$}

\end{picture}
\end{figure}

\end{document}